\documentclass[12pt,a4paper,fleqn]{elsarticle}

\usepackage{lineno}
\modulolinenumbers[5]

\usepackage{euscript,amsmath,amssymb,color}
\usepackage{rotating,float,subfigure,graphics,threeparttable}
\usepackage{natbib}
\usepackage{ulem}

\newcommand\lae[1]{\label{#1}}

\newcommand{\sub}[1] {\ensuremath{_{\text{#1}}}}

\begin{document}
\begin{frontmatter}
\title{Temperature and pressure jump coefficients at a liquid-vapor
interface}

\author[myaddress]{Denize Kalempa}

\author[sharaddress]{Irina Graur}

\address[myaddress]{Departamento de Ci\^encias B\'asicas e Ambientais, Escola de
Engenharia de Lorena, Universidade de S\~ao Paulo, 12602-810, Lorena, Brazil}

\address[sharaddress]{Aix-Marseille Universit\'e, CNRS, IUSTI UMR Marseille
7343, 13453, France}

\begin{abstract}
The temperature and pressure jump coefficients at a liquid-vapor interface are calculated from the solution of the Shakhov kinetic model for the linearized Boltzmann equation. Complete and partial evaporation/condensation at the vapor-liquid interface are assumed as the boundary condition. The discrete velocity method is used to solve the problem numerically. The jump coefficients are tabulated as functions of the evaporation/condensation coefficient. The profiles of the vapor temperature and pressure deviations from that values at the interface corresponding to the liquid temperature and saturation pressure are plotted and the solutions obtained from kinetic theory and continuum approach are shown to underline the effect of the jumps at the interface. The obtained results have been compared to those given by other authors, who applied the linearized Boltzmann equation as well as the model proposed by Bhatnagar, Gross and Krook to it, and it was found that the pressure and temperature jump coefficients are relatively insensitive to the collision laws. 

\end{abstract}


\end{frontmatter}
\section{Introduction}

In kinetic theory of gases, the problem of evaporation and condensation of a vapor on its condensed phase has been studied by many authors over the years, e.g. Refs.\cite{Pao02,Pao03,Sie04,Cip01,Son27,Loyalka:1991a}. 

As is known, from the kinetic viewpoint, during evaporation and condensation processes the vapor in the thin layer near the interface, called as Knudsen layer and whose thickness has the magnitude of a few molecular mean free paths, is in a non-equilibrium state. In the Knudsen layer, the continuum models based on the assumption of continuity of thermodynamic variables at the interface fail to predict the vapor behavior. Thus, to simulate the vapor behavior properly the methods of kinetic theory of gases based on either the solution of the Boltzmann equation and its related models, such as those proposed by Bhatnagar, Gross and Krook \cite{Bha01} and Shakhov \cite{Shk02}, referred to as BGK-model and S-model, or the Direct Simulation Monte Carlo method \cite{Bir02} must be applied. Since the methods of kinetic theory of gases require a great computational effort, for practical purposes it is still worth taking into account the gas rarefaction effects by modest computational effort and it can be done by using the special boundary conditions in the frame of the continuum approach, i.e. with the Navier-Stokes-Fourier equations.
 Thus, similar to the temperature jump on a gas-solid interface, e.g. Refs. \cite{Loy13,Son08,Sie14}, in case of a slightest gas rarefaction the continuum equations with proper boundary conditions which take into account the jumps of thermodynamic variables at the interface can be used to describe the gas behavior outside the Knudsen layer.  

In the pioneering works by Pao \cite{Pao02,Pao03} and Cipolla \textit{et al.} \cite{Cip01} the authors, by analogy with the temperature jump on a solid surface, assumed that vapor parameters could be different from that of the condensed phase and proposed the so called pressure and temperature jump boundary conditions. From the macroscopic point of view, the implementation of the jump conditions on the continuum type Navier-Stokes-Fourier equations allows to simulate correctly the vapor behavior outside the Knudsen layer. In the aforementioned papers, the collision term of the linearized Boltzmann equation was simulated via the BGK-model.
However, while Pao used the integral moment method to solve the linearized kinetic equation, which consists on obtaining a set of integral equations for the moments of the distribution function,  Cipolla \textit{et al.} used the variational method which implies the use of trial functions for the moments of the distribution function. The temperature and pressure jump coefficients arising from the conductive heat flux and mass flux were calculated by the authors and the comparison between their results obtained from different numerical techniques show a maximum difference of about 1\% for the temperature jump due to the conductive heat flux. 

The pressure and temperature jump coefficients of a gas on its plane condensed phase were also obtained by Sone and Onishi \cite{Son27} and Siewert and Thomas \cite{Sie04} from the BGK model for the linearized Boltzmann equation. However, the authors considered only the mass flux as the driving force which causes the jumps at the interface. Later, the same problem was solved by Sone \textit{et al.} \cite{Son35} with basis on the linearized Boltzmann equation. The comparison between the results obtained from the BGK kinetic model \cite{Sie04,Son27} and Boltzmann equation \cite{Son35} shows a difference of about 2\% for the temperature jump at the interface, while the difference for the density jump is negligible. 

The aim of the preset paper is to calculate the pressure and temperature jumps at the interface, caused by both heat and mass flux in the Knudsen layer, from the S-model \cite{Shk02} for the linearized Boltzmann equation. The S-model is considered the most suitable to deal with problems concerning both heat and mass transfer because it maintains the original properties of the Boltzmann equation and provides the correct Prandtl number. Complete and partial evaporation/condensation at the interface is assumed as the boundary condition. The discrete velocity method, which has been widely used to solve rarefied gas problems, is employed to solve the problem numerically. Explicit expressions of the pressure and temperature jumps coefficients related to the evaporation and heat fluxes through the interface based on the Onsager reciprocity relations are given and used as an additional criterium of the convergence of the numerical scheme. The implementation of the jump conditions in the continuum models is discussed.

\section{Statement of the problem}

Let us consider the heat and mass transfer problem in the two-phase system. 
A gas in the semi-infinite space $x' \ge 0$ is bounded by
its plane condensed phase at $x'=0$. The liquid layer has a uniform temperature
$T_0$ and saturation pressure $p_0$. The number density of saturated vapor
at $T_0$ is denoted by $n_0$. Far from the plane condensed phase, i.e. outside the Knudsen layer, the gas is
in equilibrium with number density $n_{\infty}$, temperature $T_{\infty}$, pressure $p_{\infty}=n_{\infty} k T_{\infty}$,
and flow velocity $(U_{\infty},0,0)$. 
The slow phase change is due to
small density and temperature gradients in the vapor, in the
$x'$-direction, which are given as
\begin{equation}
X_n=\biggl|\frac{\ell_0}{n_0}\frac{dn}{dx'}\biggr| \ll 1,\quad
X_T=\biggl|\frac{\ell_0}{T_0}\frac{dT}{dx'}\biggr|\ll 1,
\lae{eq1}
\end{equation} 
so the density, temperature and pressure vary on $x'$-coordinate.
The length $\ell_0$ is equivalent to the molecular mean free path of
vapor defined as 
\begin{equation}
\ell_0=\frac{\mu_0 v_0}{p_0},\quad v_0=\sqrt{\frac{2kT_0}{m}},
\lae{eq2}
\end{equation}
where $\mu_0$ is the vapor
viscosity and $v_0$ is the most probable speed of vapor molecules at the temperature $T_0$. 

Let us assume that far from the plane condensed
phase the temperature and density vary linearly on the $x'$-coordinate, while the pressure
is constant. Thus, the asymptotic behavior of the gas temperature and density read
\begin{equation}
T(x') \rightarrow T_w\left(1+\frac{x'}{\ell_0}X_T\right)\quad \text{at}\quad
\frac{x'}{\ell_0}\rightarrow \infty,
\lae{eq3}
\end{equation}
\begin{equation}
n(x') \rightarrow n_w\left(1-\frac{x'}{\ell_0}X_T\right) \quad \text{at}\quad
\frac{x'}{\ell_0}\rightarrow \infty.
\lae{eq4}
\end{equation}
Note that, since the pressure is constant far from the surface, $X_n=-X_T$ when $x'\rightarrow \infty$. $T_w$ and $n_w$ are the temperature and density of the vapor at the
plane condensed phase and their values are different from the temperature $T_0$ and density $n_0$ at the plane condensed phase. In fact, there is a
jump of both quantities at the interface so that
\begin{equation}
T_w=T_0(1+\epsilon_T),\quad n_w=n_0(1+\epsilon_n),
\lae{eq5}
\end{equation}
where $\epsilon_T$ and $\epsilon_n$ denote the macroscopic temperature and density jumps
at the interface. 

When no jump condition is assumed at the interface, the asymptotic behavior of temperature and density are denoted by
\begin{equation}
T_{NJ} (x') \rightarrow T_0\left(1+\frac{x'}{\ell_0}X_T\right) \quad \text{at}\quad
\frac{x'}{\ell_0}\rightarrow \infty,
\lae{eq05a}
\end{equation}
\begin{equation}
n_{NJ}(x')\rightarrow n_0 \left(1-\frac{x'}{\ell_0}X_T\right)\quad \text{at}\quad
\frac{x'}{\ell_0}\rightarrow \infty,
\lae{eq05b}
\end{equation}
where the subscript $NJ$ means no jump.

Thus, from (\ref{eq3}), (\ref{eq4}), (\ref{eq05a}) and (\ref{eq05b}), the macroscopic jumps can be calculated as
\begin{equation}
\epsilon_T=\lim_{x'\rightarrow
\infty}\left[\frac{T(x')-T_{NJ}(x')}{T_0}\right],\quad
\epsilon_n=\lim_{x'\rightarrow
\infty}\left[\frac{n(x')-n_{NJ}(x')}{n_0}\right].
\lae{eq5a}
\end{equation}
It is worth noting that although the jumps are used in the boundary condition, i.e. at $x'=0$, they are calculated outside the Knudsen layer. This is valid because the thickness of the Knudsen layer has the order of few
molecular mean free paths. Moreover, it is very thin compared to the characteristic size of the region occupied by the vapor. 

In order to calculate the jumps given in (\ref{eq5a}), the temperature $T(x')$ and the density $n(x')$ must be calculated from methods of kinetic theory of gases. 

The vapor pressure at the interface also experiences a jump so that 
\begin{equation}
p_w=p_0(1+\epsilon_p),
\lae{eq5b}
\end{equation}
where $\epsilon_p$ is the pressure jump. Therefore, from the ideal gas law, $p_w$=$n_wkT_w$, the pressure, temperature and number density jumps are related as
\begin{equation}
\epsilon_p=\epsilon_n+\epsilon_T.
\lae{eq5c}
\end{equation}

Here, we are going to present the values for $\epsilon_p$ and $\epsilon_T$ because, in practice, these jumps can be measured. 

\section{Kinetic equation}

For further derivation we will use the Boltzmann kinetic equation with the Shakhov model \cite{Shk02} for the collisional term. For convenience, let us introduce the dimensionless $x$-coordinate and molecular velocity as 
\begin{equation}
x=\frac{x'}{\ell_0},\quad {\bf c}=\frac{\bf v}{v_0},
\lae{k1}
\end{equation}
where $\ell_0$ and $v_0$ are defined in (\ref{eq2}).

Since we are going to calculate the pressure, temperature and density jumps with basis on macroscopic quantities outside the Knudsen layer,  two driving forces are introduced as
\begin{equation}
X_T=\frac{1}{T_0}\left(\frac{dT}{dx}\right)_{\infty},\quad X_u=\frac{U_{\infty}}{v_0},
\lae{k1a}
\end{equation}
which are assumed as being very small, i.e. 
\begin{equation}
|X_T|\ll 1, \quad |X_u|\ll 1.
\lae{k2}
\end{equation}
Note that, outside the Knudsen layer the pressure is constant and, as a consequence, the thermodynamic forces given in (\ref{eq1}) are related as $X_T$=$-X_n$.
The assumption of smallness of the driving forces allows us to solve the problem 
on the basis of a linearized approach. Thus, the distribution function of molecular
velocities can be represented as
\begin{equation}
f(x,{\bf c})=f_0^M[1+h^{(T)}(x,{\bf c})X_T+h^{(u)}(x,{\bf c})X_u],
\lae{k3}
\end{equation}
where 
\begin{equation}
f_0^M=\frac{n_0}{(\sqrt{\pi}v_0)^3}\mbox{e}^{-c^2}
\lae{k4}
\end{equation}
is the Maxwellian function with the plane condensed phase characteristics, 
and $h^{(i)}(x,{\bf c})$ ($i$=$T$,$u$) is the perturbation function due
to the corresponding thermodynamic force.   

Thus, in the absence of
external forces, after the substitution of the representation (\ref{k3}) into the Shakhov model kinetic equation \cite{Shk02}, two independent kinetic equations are obtained for the problem in question as
\begin{equation}
c_x\frac{\partial h^{(i)}}{\partial x}=\hat{L}h^{(i)},\quad i=T, u,
\lae{k5}
\end{equation} 
where the collisional operator reads
\begin{equation}
\hat{L}h^{(i)}=\nu^{(i)} + \left(c^2-\frac 32 \right)\tau^{(i)} +
2c_xu_x^{(i)}+\frac{4}{15}c_x\left(c^2-\frac 52 \right)q_x^{(i)} - h^{(i)}.
\lae{k6}
\end{equation}
The dimensionless quantities $\nu^{(i)}$,  
$\tau^{(i)}$, $u_x^{(i)}$ and $q_x^{(i)}$  appearing in (\ref{k6}) correspond to the density and temperature
deviations from the values $n_0$ and $T_0$ at the plane condensed phase, bulk velocity and heat flux in the $x$-direction, respectively, caused by the
corresponding driving force. According to Refs. \cite{Fer02,Cha04}, in kinetic theory of gases the macroscopic quantities which characterize the gas flow are calculated in terms of the distribution function of molecular velocities. In our notation, the representation (\ref{k3}) allows to write the density and temperature deviations as  
\begin{equation}
\nu(x)=
\frac{n(x)-n_0}{n_0}=\nu^{(T)}(x)X_T + \nu^{(u)}(x)X_u,
\lae{f1}
\end{equation}
\begin{equation}
\tau(x)=
\frac{T(x)-T_0}{T_0}=\tau^{(T)}(x)X_T + \tau^{(u)}(x)X_u,
\lae{f2}
\end{equation}
where 
\begin{equation}
\nu^{(i)}(x)=\frac{1}{\pi^{3/2}}\int
h^{(i)}(x,{\bf{c}})\mbox{e}^{-c^2}\, \mbox{d}{\bf c},
\lae{k7}
\end{equation}
\begin{equation}
\tau^{(i)}(x)=\frac{2}{3\pi^{3/2}}\int
h^{(i)}(x,{\bf{c}})\left(c^2-\frac 32\right)\mbox{e}^{-c^2}\, \mbox{d}{\bf c}.
\lae{k8}
\end{equation}
Moreover, the bulk velocity and heat flux read
\begin{equation}
u(x)=
\frac{U_x(x)}{v_0}=u_x^{(T)}(x)X_T + u_x^{(u)}(x)X_u,
\lae{f3}
\end{equation}
\begin{equation}
q(x)=
\frac{Q_x(x)}{v_0p_0}=q_x^{(T)}(x)X_T + q_x^{(u)}(x)X_u,
\lae{f4}
\end{equation}
where 
\begin{equation}
u_x^{(i)}(x)=\frac{1}{\pi^{3/2}}\int
h^{(i)}(x,{\bf{c}})c_x\mbox{e}^{-c^2}\, \mbox{d}{\bf c},
\lae{k9}
\end{equation}
\begin{equation}
q_x^{(i)}(x)=\frac{1}{\pi^{3/2}}\int
h^{(i)}(x,{\bf{c}})\left(c^2-\frac 52\right)c_x\mbox{e}^{-c^2}\, \mbox{d}{\bf c}.
\lae{k10}
\end{equation}

Therefore, from (\ref{eq5a}) one can easily see that the temperature and density jumps can be calculated in terms of the density and temperature deviations defined in (\ref{k7}) and (\ref{k8}) as 
\begin{equation}
\epsilon_T=\lim_{x\rightarrow \infty}\left[(\tau^{(T)}-x)X_T+\tau^{(u)}X_u\right],
\lae{k10a0}
\end{equation}
\begin{equation}
\epsilon_n=\lim_{x\rightarrow \infty}\left[(\nu^{(T)}+x)X_T+\nu^{(u)}X_u\right].
\lae{k10a1}
\end{equation}

\section{Boundary condition}

It is assumed that a fraction $\sigma$ of
incident molecules condensates at the plane  condensed phase and then evaporates from it with the Maxwellian distribution function $f_0^M$ given in (\ref{k4}). Meanwhile, the fraction $1-\sigma$ is reflected from the condensed phase under the assumption of diffuse scattering. Therefore, the disbribution function of emmited molecules from the condensed phase is written as
\begin{equation}
f(0,{\bf c})=\sigma f_0^M + (1-\sigma)f_r\quad
\text{at} \quad c_x>0,
    \lae{b1}
\end{equation}
where 
\begin{equation}
f_r=\frac{n_r}{n_0}f_0^M
    \lae{b2}
\end{equation}
is the distribution function of the molecules diffusely reflected from the condensed phase. The constant $n_r$ is the number density calculated from the impermeability condition for the molecules diffusely reflected from the plane condensed phase. Since the representation (\ref{k3}) is used to linearize the Shakhov kinetic equation, the boundary condition (\ref{b1}) to solve the equations given in (\ref{k5}) reads
\begin{equation}
h^{(i)}(0,{\bf c})=-(1-\sigma)\xi^{(i)}\quad \text{at}\quad c_ x > 0,
\lae{k10a}
\end{equation}
where
\begin{equation}
\xi^{(i)}=1-\frac{n_r^{(i)}}{n_0}=\frac{2}{\pi}\int_{c_x< 0}c_xh^{(i)}(0,{\bf c})\mbox{e}^{-c^2}\,
\mbox{d}{\bf c}.
\lae{k10b}
\end{equation}
It is worth underling that in the case of slow evaporation and condensation these processes are symmetric, i.e. the results can be obtained just by changing the sign of the evaporation (condensation) velocity.
\section{Condition far from the condensed phase ($x \rightarrow \infty$)}

Far from the plane condensed phase, the solution of the problem is obtained from the Chapman-Enskog approach to the kinetic equation. Thus, the distribution function of molecular velocities can be written as
\begin{equation}
f=f_{\infty}^M(1+h_{CE}^{(T)}X_T + h_{CE}^{(u)}X_u)
\lae{ce1}
\end{equation}
where the Maxwellian function corresponding to the equilibrium reads
\begin{equation}
f_{\infty}^M=n_{\infty}\left(\frac{m}{2\pi kT_{\infty}}\right)^{3/2}
\exp{\left[-\frac{m}{2kT_{\infty}}({\bf v}-{\bf
U}_{\infty})^2\right]},
\lae{k10c}
\end{equation}
with $T_{\infty}$ and $n_{\infty}$ given in (\ref{eq3}) and
(\ref{eq4}), while ${\bf U}_{\infty}=(U_{\infty},0,0)$. This function is related to the distribution function $f^M_0$, defined in
(\ref{k4}), as follows
\begin{equation}
f_{\infty}^M=f_0^M\left[1+\epsilon_n+\left(c^2-\frac
32\right)\epsilon_T+x\left(c^2-\frac 52\right)X_T+2c_xX_u\right].
\lae{k10d}
\end{equation}

The use of the representation (\ref{ce1}) into the Shakhov kinetic equation allows to find the Chapman-Enskog solution as
\begin{equation}
h_{CE}^{(T)}=-\frac 32 c_x\left(c^2-\frac 52 \right),\quad h_{CE}^{(u)}=0.
\lae{k15}
\end{equation}
Note that, from (\ref{k10a0}) and (\ref{k10a1}), the temperature and density
jumps can be written as
\begin{equation}
\epsilon_T=\epsilon_T^{(T)}X_T + \epsilon_T^{(u)}X_u,
\lae{k11}
\end{equation}
\begin{equation}
\epsilon_n=\epsilon_n^{(T)}X_T + \epsilon_n^{(u)}X_u,
\lae{k12}
\end{equation}
where the jump coefficients are defined as
\begin{equation}
\epsilon_T^{(T)}=\lim_{x\rightarrow \infty} (\tau^{(T)}-x),\quad 
\epsilon_T^{(u)}=\lim_{x\rightarrow \infty} \tau^{(u)},
\lae{k12a}
\end{equation}
\begin{equation}
\epsilon_n^{(T)}=\lim_{x\rightarrow \infty} (\nu^{(T)}+x),\quad 
\epsilon_n^{(u)}=\lim_{x\rightarrow \infty} \nu^{(u)}.
\lae{k12b}
\end{equation}

Thus, after substituting the representation (\ref{k11}) and  (\ref{k12})
into (\ref{k10d}), the following asymptotic behaviors of the perturbation functions are obtained from (\ref{ce1}) as 
\begin{equation}
h^{(T)}(x,{\bf c})=\epsilon_n^{(T)}+\left(c^2-\frac
32\right)\epsilon_T^{(T)}+x\left(c^2-\frac 52 \right) + h_{CE}^{(T)}\quad
\text{at}\quad x\rightarrow \infty,
\lae{k13}
\end{equation}
\begin{equation}
h^{(u)}(x,{\bf c})=\epsilon_n^{(u)}+\left(c^2-\frac
32\right)\epsilon_T^{(u)}+ 2c_x +h_{CE}^{(u)}\quad
\text{at}\quad x\rightarrow \infty.
\lae{k14}
\end{equation}
where $h_{CE}^{(T)}$ and $h_{CE}^{(u)}$ are given in (\ref{k15}). 

Note that, from the relation (\ref{eq5c}), 
\begin{equation}
\epsilon_n^{(i)}=\epsilon_{p}^{(i)}-\epsilon_{T}^{(i)},\quad i=u, T.
\end{equation}


\section{Numerical solution}

The linearized kinetic equations given in (\ref{k5}) subject to the corresponding conditions given in (\ref{k10a}), (\ref{k13}) and (\ref{k14}), are solved numerically
by using the discrete velocity method whose details can
be found in the literature, see e.g. Ref. \cite{Sha02B}. Firstly, to reduce the
number of variables in the molecular velocity space and, consequently, to reduce the computational effort to solve the problem, new functions are introduced as
\begin{equation}
\phi^{(i)}(x, c_x)=\frac{1}{\pi}\int \int h^{(i)}(x,{\bf
c})\mbox{e}^{-c_y^2-c_z^2}\, \mbox{d}c_y\mbox{d}c_z,
\lae{k18}
\end{equation}
\begin{equation}
\psi^{(i)}(x, c_x)=\frac{1}{\pi}\int \int h^{(i)}(x,{\bf
c})\mbox{e}^{-c_y^2-c_z^2}(c_y^2+c_z^2-1)\, \mbox{d}c_y\mbox{d}c_z.
\lae{k19}
\end{equation}
The substitution of these new functions into (\ref{k5}) leads to the following system of kinetic equations for each thermodynamic force 
\begin{equation}
c_x\frac{\partial \phi^{(i)}}{\partial x}=\hat{L}\phi^{(i)},
\lae{k20}
\end{equation}
\begin{equation}
c_x\frac{\partial \psi^{(i)}}{\partial x}=\hat{L}\psi^{(i)},
\lae{k21}
\end{equation}
where
\begin{equation}
\hat{L}\phi^{(i)}=\nu^{(i)}+\left(c_x^2-\frac
12\right)\tau^{(i)}+2c_xu_x^{(i)}+\frac{4}{15}c_x\left(c_x^2-\frac
32\right)q_x^{(i)}-\phi^{(i)},
\lae{k22}
\end{equation}
\begin{equation}
\hat{L}\psi^{(i)}=\tau^{(i)}+\frac{4}{15}c_xq_x^{(i)}-\psi^{(i)}.
\lae{k23}
\end{equation}

The moments of the perturbation function given in (\ref{k7})-(\ref{k10}) are
written in terms of the new functions as
\begin{equation}
\nu^{(i)}(x,c_x)=\frac{1}{\sqrt{\pi}}\int \phi^{(i)}\mbox{e}^{-c_x^2}\,
\mbox{d}c_x,
\lae{k24}
\end{equation}
\begin{equation}
\tau^{(i)}(x,c_x)=\frac{2}{3\sqrt{\pi}}\int \left[\left(c_x^2-\frac
12\right)\phi^{(i)}+\psi^{(i)}\right]\mbox{e}^{-c_x^2}\,
\mbox{d}c_x,
\lae{k25}
\end{equation}
\begin{equation}
u_x^{(i)}(x,c_x)=\frac{1}{\sqrt{\pi}}\int \phi^{(i)}c_x\mbox{e}^{-c_x^2}\,
\mbox{d}c_x,
\lae{k26}
\end{equation}
\begin{equation}
q_x^{(i)}(x,c_x)=\frac{1}{\sqrt{\pi}}\int \left[\left(c_x^2-\frac
32\right)\phi^{(i)}+\psi^{(i)}\right]c_x\mbox{e}^{-c_x^2}\,
\mbox{d}c_x.
\lae{k27}
\end{equation}

Moreover, from (\ref{k10a}), the boundary conditions at $x$=0  are written in terms of the new functions as
\begin{equation}
\phi^{(i)}(0,c_x)=-(1-\sigma)\xi^{(i)},\quad c_x >0
\lae{k28}
\end{equation}
\begin{equation}
\psi^{(i)}(0,c_x)=0,\quad c_x>0,
\lae{k29}
\end{equation}
where 
\begin{equation}
\xi^{(i)}=2\int_{c_x<0} \phi^{(i)}(0,c_x) c_x\mbox{e}^{-c_x^2}\, \mbox{d}c_x.
\lae{ke30}
\end{equation}
Note that, the boundary condition depends on $\xi^{(i)}$ only when $\sigma
\ne 1$, which corresponds to the partial, or non-complete, condensation-evaporation at the interface. 

Far from the boundary, i.e. in the limit $x\rightarrow \infty$, the
conditions (\ref{k13}) and (\ref{k14}) are written in terms of the functions as
\begin{equation}
\phi^{(T)}(x,c_x)=\epsilon_n^{(T)}+\left(c_x^2-\frac
12\right)\epsilon_T^{(T)}+x\left(c_x^2-\frac 32\right)-\frac 32
c_x\left(c_x^2-\frac 32\right),
\lae{k30}
\end{equation}
\begin{equation}
\psi^{(T)}(x,c_x)=\epsilon_T^{(T)}+x-\frac 32c_x,
\lae{k31}
\end{equation}
\begin{equation}
\phi^{(u)}(x,c_x)=\epsilon_n^{(u)}+\left(c_x^2-\frac
12\right)\epsilon_T^{(u)}+2c_x,
\lae{k32}
\end{equation}
\begin{equation}
\psi^{(u)}(x,c_x)=\epsilon_T^{(u)}.
\lae{k33}
\end{equation}

Thus, for each thermodynamic force, the system of kinetic equations given by 
(\ref{k20}) and (\ref{k21}) with
the corresponding boundary condition and asymptotic behavior was solved
numerically via the discrete velocity method with an numerical error of 0.1\% for the temperature and density jump coefficients at the vapor-liquid interface. The
Gaussian-Hermite quadrature was employed to discretize the $x$-component of
the molecular velocity space and calculate the macroscopic characteristics of the gas
flow corresponding the the density and temperature deviations, bulk velocity
and heat flux. Details concerning the numerical technique to found the nodes
and weights can be found in Ref. \cite{kry02}. A finite difference scheme
was used to calculate the derivatives in the kinetic equations. The accuracy
was estimated by varying the grid parameters $N_x$ and $N_c$, corresponding to
the number of nodes in the $x$-coordinate and molecular velocity component
$c_x$, as well and the maximum distance $x_{max}$ from the interface. The
value of $N_c$ was fixed at 12, while the value of $N_r$ varied according to
$x_{max}$ so that the increment $\Delta x \sim 10^{-3}$. 

\section{Thermodynamic analysis}

As it is known from the thermodynamic of irreversible processes, e.g.
Ref.\cite{DeG01}, in a weakly disturbed system, the thermodynamic fluxes are linearly related to thermodynamic forces. In the present problem, the linear relations read
\begin{equation}
J_M'=\Lambda_{11}'X_p + \Lambda_{12}'X_T,
\lae{k17b}
\end{equation}
\begin{equation}
J_T'=\Lambda_{21}'X_p + \Lambda_{22}'X_T,
\lae{k17c}
\end{equation}
where the thermodynamic fluxes conjugated to the pressure $X_p$ and temperature $X_T$ gradients are introduced as 
\begin{equation}
J_M'=n_0U_x=n_0v_0u,\quad 
J_T'=\frac{Q_x}{kT_0}=n_0v_0q.
\lae{k17a}
\end{equation}
The dimensionless quantities $u$ and $q$ are given in (\ref{f3}) and
(\ref{f4}), and outside of the Knudsen layer both quantities are constant. $\Lambda_{ij}'$ ($i,j$=1, 2) are referred to as kinetic coefficients which satisfy the Onsager reciprocity relation $\Lambda_{12}'$=$\Lambda_{21}'$.

Moreover, inverted linear relations can be obtained from (\ref{k17b}) and (\ref{k17c}), so that the thermodynamic forces can be
written in terms of the thermodynamic fluxes as
\begin{equation}
X_p=a_{11}'J_M' + a_{12}'J_T',
\lae{k17d}
\end{equation}
\begin{equation}
X_T=a_{21}'J_M'+a_{22}'J_T',
\lae{k17e}
\end{equation}
where the new kinetic coefficients read
\begin{equation}
a_{11}'=\frac{\Lambda_{22}'}{\EuScript{D}},\quad
a_{12}'=a_{21}'=-\frac{\Lambda_{12}'}{\EuScript{D}},\quad
a_{22}'=\frac{\Lambda_{11}'}{\EuScript{D}},
\lae{k17f}
\end{equation}
and
\begin{equation}
   \EuScript{D}=\Lambda_{11}'\Lambda_{22}'-\Lambda_{12}'\Lambda_{21}'. 
\end{equation}

In our approach, the jump coefficients are calculated in terms of
macroscopic quantities outside the Knudsen layer. However, in the macroscopic point of view, we are extrapolating the continuum profiles of thermodynamic variables through the Knudsen layer to the interface. Then, the pressure and temperature jump coefficients can also be written as
\begin{equation}
\epsilon_p=\frac{p_w-p_0}{p_0},\quad \epsilon_T=\frac{T_w-T_0}{T_0}.
\lae{ext1}
\end{equation}
Thus, after some algebraic manipulation, the representations (\ref{k11}) and (\ref{k12}) allow to obtain relations similar to those given in (\ref{k17d}) and (\ref{k17e}) as 
\begin{equation}
\frac{p_w-p_0}{p_0}=\epsilon_p^{(u)}\frac{J_M'}{n_0v_0}-\frac{8}{15}\epsilon_p^{(T)}\frac{J_T'}{n_0v_0},
\lae{ext2}
\end{equation}
\begin{equation}
\frac{T_w-T_0}{T_0}=\epsilon_T^{(u)}\frac{J_M'}{n_0v_0}-\frac{8}{15}\epsilon_T^{(T)}\frac{J_T'}{n_0v_0},
\lae{ext3}
\end{equation}
where the Fourier law was used to write the conductive heat flux as $q$=$-15X_T/8$. These relations allow us to verify the fullfillment of the reciprocity relation
\begin{equation}
\epsilon_T^{(u)}=-\frac{8}{15}\epsilon_p^{(T)}.
\lae{ext4}
\end{equation}

The fullfillment of the reciprocity relation (\ref{ext4}) is verified numerically as an additional criterium of the convergence of the numerical scheme.

\section{Results}

Table \ref{tab1} presents the pressure and temperature jump coefficients obtained
in the present work under the assumption of complete evaporation and condensation at the interface. The results on the jumps provided  by Pao \cite{Pao02} and  Cipolla \textit{et al.} \cite{Cip01} due to both heat and mass transfer are given in Table \ref{tab1} for comparison. In both papers the BGK kinetic model was used to solve the problem, but the numerical techniques are different. Pao used the integral moment method, which consists on obtaining a set of integral equations for the moments of the distribution function, 
while Cipolla \textit{et al.} used the variational method which implies the use of trial functions for the moments of the distribution function. The results for the jumps due to only the mass flux obtained  by Sone and Onishi \cite{Son27} from the BGK kinetic model and those obtained by Sone \textit{et al.} \cite{Son35} from the Boltzmann equation with the original collision term are also given in Table \ref{tab1}. It is worth noting that, in Ref. \cite{Son35} the rigid-spheres intermolecular model was used to solve the Boltzmann equation. Loyalka  \cite{Loyalka:1991a} analyzed the influence of intermolecular force laws on the values of the jump coefficients and he concluded that these coefficients are relatively insensitive to them.

According to the tabulated results, there is a good agreement between the present results and those given in Ref. \cite{Cip01}. The maximum difference between the results is approximately equal to 1.2\% for the coefficient $\epsilon_T^{(T)}$. However, it is interesting to note that the coefficient $\epsilon_{T}^{(T)}$ which arises due to the conductive heat flux corresponds to the temperature jump coefficient calculated in case of diffuse scattering of gas molecules from a solid surface. According to the literature review on slip and temperature jumps \cite{Sha84}, for practical applications the value of 1.95 is recommended for this jump coefficient. Concerning the
comparison with the results given in Refs. \cite{Pao02,Son27}, the maximum difference is negligible. Thus, Table \ref{tab1} shows 
that even if the collision terms of the BGK and S models are different both kinetic models provide the same results for the jump coefficients. The small difference in the comparison with the results given by Cipolla \textit{et al.} is due to the use of trial functions in the variational method which introduces numerical error. 

The comparison with the results obtained from the Boltzmann equation \cite{Son35} shows a difference of about 0.5\% for the pressure jump and 2\% for the temperature jump due to the mass flux in the Knudsen layer. Thus, since to solve the Boltzmann equation is still a difficult task which demands a great computational effort, the use of kinetic models plays an important role in the solution of practical problems. 

It is worth noting that the Shakhov model is the most suitable for solving problems concerning both heat and mass transfer. The collision term depends on the collision frequency \textit{f}. In the S-model the collision frequency is given by the ratio between the pressure and viscosity of the gas, i.e. $f=p/\mu$. When the BGK model is used with the same collision frequency, the results for $\epsilon_p^{(u)}$ and $\epsilon_T^{(u)}$ are equal to those obtained from the S-model. Nonetheless, the results for $\epsilon_p^{(T)}$ and $\epsilon_T^{(T)}$ must be multiplied by 3/2 due to the fact that the BGK model provides the correct Prandt number when the collision frequency is chosen as $f=2p/(3\mu)$.

From the provided analysis we may conclude that the values of jump coefficients are less sensitive not only to the intermolecular force laws as it was found in Ref. \cite{Loyalka:1991a}, but also to the collisional laws.

The results for partial or non-complete evaporation-condensation at the interface are provided in Table \ref{tab2}. According to these results, only the coefficients $\epsilon_p^{(T)}$ and $\epsilon_p^{(u)}$ are impacted by the non-complete evaporation-condensation condition. The smaller the evaporation/condensation coefficient $\sigma$ the larger the magnitude of these jump coefficients.  The maximum deviation of $\epsilon_p^{(T)}$ from the corresponding value for complete evaporation/condensation is less than 0.5\% so that the influence of $\sigma$ in such jump coefficient is negligible. Meanwhile, the deviation of $\epsilon_p^{(u)}$ from that value for complete evaporation/condensation at the interface is larger than 100\% when $\sigma \le 0.6$. Therefore, one can say that only the jump coefficient $\epsilon_p^{(u)}$ is indeed impacted by the evaporation/condensation coefficient $\sigma$. A similar conclusion is given by Cipolla \textit{et al.} \cite{Cip01} and their results for $\epsilon_p^{(u)}$ are shown in Table \ref{tab2} for comparison.

\begin{figure}
\centering
\includegraphics[scale=1.]{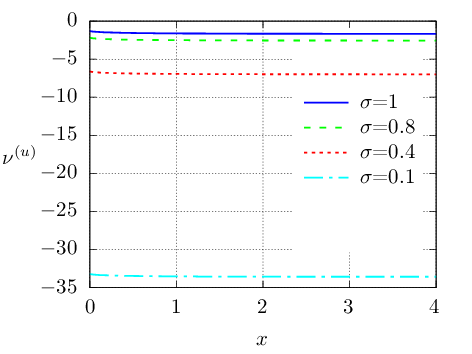}
\caption{Density deviation due to $X_u$.}\label{fig:nu}
\end{figure}

The significant dependence of $\epsilon_p^{(u)}$ on the evaporation/condensation coefficient $\sigma$ is due to the dependence of the density deviation $\nu^{(u)}$ on such coefficient. Figure \ref{fig:nu} shows the profile of $\nu^{(u)}$ as function of the $x$-coordinate for $\sigma$=0.1, 0.4, 0.8 and 1.  

In the following it is shown how the derived jump conditions modify the temperature and pressure profiles obtained from the continuum equations.

\begin{figure}
\centering
\includegraphics[scale=1.]{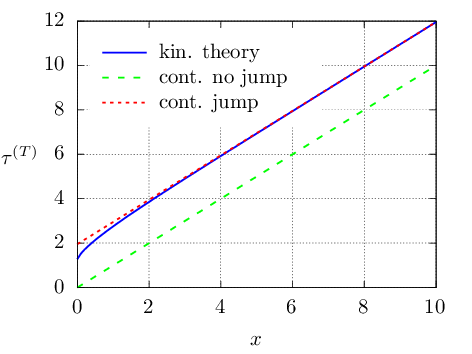}
\includegraphics[scale=1.]{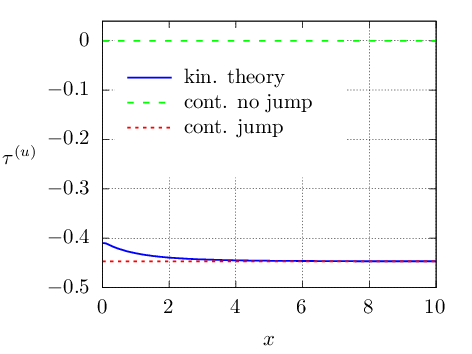}
\caption{Temperature deviation due to $X_T$ and $X_u$.}\label{fig:T_jump}
\end{figure}

In the case of continuum solution, no temperature jump is assumed on the interface and the vapor temperature at the interface, $T_w$, is equal to the temperature of the interface, i.e. $T_w=T_0$. In this case, the temperature deviation given in (\ref{f2}) is obtained from (\ref{eq05a}) as
\begin{equation}\label{eq:tau_cont}
\tau(x)=x X_T,
\end{equation}
so that
\begin{equation}\label{eq:tau_Tu_cont}
\tau^{(T)}(x)=x \qquad \rm{and} \qquad 
\tau^{(u)}=0.
\end{equation}
Figure  \ref{fig:T_jump} shows the temperature  deviations given in (\ref{eq:tau_Tu_cont}) by green lines.

When the temperature jump condition is used, the temperature deviation obtained in the frame of continuum approach with the vapor temperature at the interface, $T_w$, given in (\ref{eq5}) reads
\begin{equation}\label{eq:tau_cont_jump}
\tau(x)=(\epsilon^{(T)}_T+ x) X_T+\epsilon^{(u)}_T X_u,
\end{equation}
where
\begin{equation}\label{eq:tau_Tu_jump}
\tau^{(T)}(x)=\epsilon^{(T)}_T+x \qquad \text{and} \qquad 
\tau^{(u)}(x)=\epsilon^{(u)}_T.
\end{equation}
These temperature deviations are also shown in Figure  \ref{fig:T_jump} by red lines. 

The temperature deviations calculated numerically from the solution of the Shakhov kinetic equation are also shown in Figure  \ref{fig:T_jump} by blue lines. As one can see from Figure  \ref{fig:T_jump}, outside the Knudsen layer the use of the temperature jump boundary condition leads to the same solution obtained from kinetic theory. Inside the Knudsen layer the profile obtained from kinetic theory is different from that obtained in the frame of continuum approach, but when rarefaction is small such an  effect is neglected because the thickness of the Knudsen layer.

Similarly, the pressure deviation can also be plotted for comparison. Let us write the pressure deviation as
\begin{equation}
\eta(x)=\frac{p(x)-p_0}{p_0}=\eta^{(T)}X_T + \eta^{(u)}X_u.
\label{press-dev}
\end{equation}

\begin{figure}
\centering
\includegraphics[scale=1.]{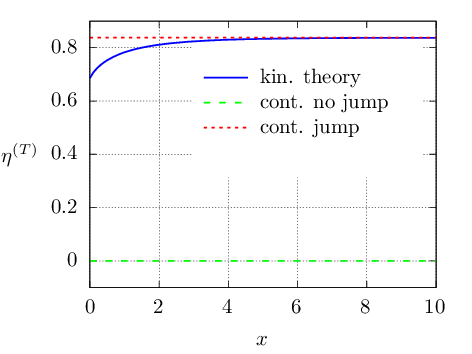}
\includegraphics[scale=1.]{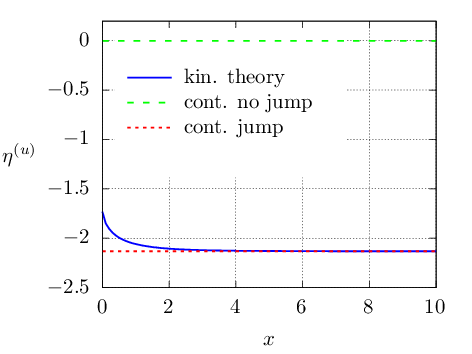}
\caption{Pressure deviation due to $X_T$ and $X_u$.}\label{fig:P_jump}
\end{figure}

When no jump condition is assumed so that $p_w=p_0$, the profiles given in (\ref{eq05a}) and (\ref{eq05b}) allow us to obtain the pressure deviations as 
\begin{equation}
\eta^{(T)}(x)=0,\qquad \eta^{(u)}(x)=0.
\label{press-dev1}
\end{equation}
Otherwise, when the pressure jump condition is assumed, the vapor pressure at the interface, $p_w$, is given in (\ref{eq5b}). Thus, the pressure deviations obtained from (\ref{eq3}) and (\ref{eq4}) read
\begin{equation}
\eta^{(T)}(x)=\epsilon_p^{(T)},\qquad \eta^{(u)}(x)=\epsilon_p^{(u)}.
\label{press-dev2}
\end{equation}
Figure \ref{fig:P_jump} shows the pressure deviations profiles given in (\ref{press-dev1}) and (\ref{press-dev2}) by green and red lines, respectively. The pressure deviations obtained numerically from the Shakhov kinetic equation are also plotted for comparison.


\section{Application to evaporation problem}

The derived expressions (\ref{ext2}) and (\ref{ext3}) allow to relate the evaporation and heat fluxes through an interface to the pressure and temperature jumps. We can write them in another form as 
\begin{eqnarray}\label{eq:BC_jump_p}
\frac{p\sub{v}-p\sub{sat}(T\sub{L})}{p\sub{sat}(T\sub{L})}=\epsilon^{(u)}_p\frac{J}{\rho\sub{sat}(T\sub{L}) \sqrt{2{\cal R}T\sub{L}}}-\frac{8}{15}\epsilon^{(T)}_p\frac{q\sub{v}}{p\sub{sat}(T\sub{L})\sqrt{2{\cal R}T\sub{L}}},
\end{eqnarray} 
\begin{eqnarray}\label{eq:BC_jump_T}
\frac{T\sub{v}-T\sub{L}}{T\sub{L}}=\epsilon^{(u)}_T\frac{J}{\rho\sub{sat}(T\sub{L}) \sqrt{2{\cal R}T\sub{L}}}+-\frac{8}{15}\epsilon^{(T)}_T\frac{q\sub{v}}{p\sub{sat}(T\sub{L}) \sqrt{2{\cal R}T\sub{L}}}.
\end{eqnarray} 
Here $T\sub{L}$ is the liquid-vapor interface temperature from the liquid side, $p\sub{sat}(T\sub{L})$ is the saturation pressure at this temperature, 
$\rho\sub{sat}(T\sub{L})$ is the saturation density, $p\sub{v}$ is the vapor pressure,  $T\sub{v}$ is the liquid-vapor interface temperature from the vapor side, ${\cal R}$ is the gas-specific constant. The specific quantities such as the saturation pressure and density for a given liquid temperature can be found in handbooks for a given liquid nature. 
As during the evaporation or condensation a vapor near a liquid interface is in its non-equilibrium state these conditions must be applied when simulating the evaporation and condensation using the continuum approach.

One example of application of these conditions to the FC-72 liquid evaporation can be found in Ref. \cite{Graur:2024}. The results obtained by using the jump conditions (\ref{eq:BC_jump_p})-(\ref{eq:BC_jump_T}) are compared to that obtained from the widely used Schrage expression.

\section{Conclusion}
The temperature and pressure jump coefficients at the liquid vapor interface are calculated from the numerical solution of the Shakhov model kinetic equation. The cases of complete and partial evaporation and condensation processes are considered. The numerical values of the obtained coefficients are very close to those available in the literature, obtained by using another approaches such as the variational method and the integral moment method to solve the BGK kinetic model. The comparison with results obtained from the Boltzmann equation shows a good agreement for the jump coefficients due the mass flux in the Knudsen layer. Therefore, one may conclude that the jump coefficients are less sensitive not only to the intermolecular force laws as it was found previously, but also to the collisional laws. For practical purposes, the Shakhov model is the most suitable for solving problems concerning both heat and mass transfer. Moreover, its solution via the discrete velocity method is more attractive because it provides reliable results with less computational effort compared to that required when other numerical approaches are used.  

\section*{Acknowledgments}

The authors acknowledge the Funda\c{c}\~ao de Pesquisa do Estado de S\~ao
Paulo (FAPESP, Brazil), grant 2022/10476-1, for the support of the research.

%
\bibliographystyle{unsrt}

\begin{thebibliography}{10}

\bibitem{Pao02}
Y-P Pao.
\newblock Temperature and density jumps in the kinetic theory of gases and
  vapors.
\newblock {\em Phys. Fluids}, 14(7):1340--1346, 1971.

\bibitem{Pao03}
Y-P Pao.
\newblock Application of kinetic theory to the problem of evaporation and
  condensation.
\newblock {\em Phys. Fluids}, 14(2):306--312, 1971.

\bibitem{Sie04}
C~E Siewert and J~R Thomas.
\newblock Half-space problems in the kinetic theory of gases.
\newblock {\em Phys. Fluids}, 16:1557--1559, 1973.

\bibitem{Cip01}
J~W Cipolla~Jr, H~Lang, and S~K Loyalka.
\newblock Kinetic theory of condensation and evaporation. {II}.
\newblock {\em J. Chem. Phys.}, 61:69--77, 1974.

\bibitem{Son27}
Y~Sone and Y~Onishi.
\newblock Kinetic theory of evaporation and condensation.
\newblock {\em J. Phys. Soc. Jpn.}, 35:1773--1776, 1973.

\bibitem{Loyalka:1991a}
S.~K. Loyalka.
\newblock Kinetic theory of planar condensation and evaporation.
\newblock {\em Transport theory and statistical physics}, 20(2-3):237--249,
  1991.

\bibitem{Bha01}
P~L Bhatnagar, E~P Gross, and M~A Krook.
\newblock A model for collision processes in gases.
\newblock {\em Phys. Rev.}, 94:511--525, 1954.

\bibitem{Shk02}
E~M Shakhov.
\newblock Generalization of the {K}rook kinetic relaxation equation.
\newblock {\em Fluid Dynamics}, 3(5):95--96, 1968.

\bibitem{Bir02}
G~A Bird.
\newblock {\em Molecular Gas Dynamics and the Direct Simulation of Gas Flows}.
\newblock Oxford University Press, Oxford, 1994.

\bibitem{Loy13}
S~K Loyalka.
\newblock Temperature jump and thermal creep slip: Rigid sphere gas.
\newblock {\em Phys. Fluids A}, 1:403--408, 1989.

\bibitem{Son08}
Y~Sone, T~Ohwada, and K~Aoki.
\newblock Temperature jump and {K}nudsen layer in a rarefied gas over a plane
  wall: Numerical analysis of the linearized {B}oltzmann equation for
  hard-sphere molecules.
\newblock {\em Phys. Fluids A}, 1(2):363--370, 1989.

\bibitem{Sie14}
C~E Siewert.
\newblock The linearized {B}oltzmann equation: {A} concise and accurate
  solution of the temperature-jump problem.
\newblock {\em J. Quant. Spec. Rad. Tran}, 77:417--432, 2003.

\bibitem{Son35}
Y~Sone, T~Ohwada, and K~Aoki.
\newblock Evaporation and condensation on a plane condensed phase: numerical
  analysis of the linearized {B}oltzmann equation for hard-sphere molecules.
\newblock {\em Phys. Fluids A}, 1:1398--1405, 1989.

\bibitem{Fer02}
J~H Ferziger and H~G Kaper.
\newblock {\em Mathematical Theory of Transport Processes in Gases}.
\newblock North-Holland Publishing Company, Amsterdam, 1972.

\bibitem{Cha04}
S~Chapman and T~G Cowling.
\newblock {\em The Mathematical Theory of Non-Uniform Gases}.
\newblock University Press, Cambridge, 3 edition, 1970.

\bibitem{Sha02B}
F~Sharipov.
\newblock {\em Rarefied Gas Dynamics. Fundamentals for Research and Practice}.
\newblock Wiley-VCH, Berlin, 2016.

\bibitem{kry02}
V~I Krylov.
\newblock {\em Approximate Calculation of Integrals}.
\newblock Dover Publication Inc., Mineola, 2005.

\bibitem{DeG01}
S~R De~Groot and P~Mazur.
\newblock {\em Non-Equilibrium Thermodynamics}.
\newblock Dover Publications, Inc., New York, 1984.

\bibitem{Sha84}
F~Sharipov.
\newblock Data on the velocity slip and temperature jump on a gas-solid
  interface.
\newblock {\em J. Phys. Chem. Ref. Data}, 40(2):023101, 2011.

\bibitem{Graur:2024}
I.~Graur, A.~Rednikov, and L.~Ronshin, F.and~Tadrist.
\newblock Numerical modeling and simulation of near-contact-line evaporation
  kinetics.
\newblock In {\em Proceedings of 9th European Thermal Sciences Conference
  (Eurotherm 2024), Slovenia}, 2024.

\end{thebibliography}



\clearpage
\begin{table}
\centering
\caption{Temperature and pressure jump coefficients in the case of complete
condensation and evaporation at the interface ($\sigma$=1).}
\begin{tabular}{cccccc}\hline
 Jump  & Present & Cipolla \textit{et al.} & Pao & Sone/Onishi & Sone \textit{et al.} \\ 
Coefficient & Work & Ref.\cite{Cip01} & Ref.\cite{Pao02} & Ref. \cite{Son27} & Ref. \cite{Son35} \\ \hline
$\epsilon_T^{(T)}$ & 1.9540 & 1.9309 & 1.9540 & ----- & ----- \\
$\epsilon_P^{(T)}$ & 0.8376 & 0.8385 & 0.8376 & ----- & ----- \\
$\epsilon_T^{(u)}$ & -0.4467 & -0.4472 & -0.4468 & -0.4467 & -0.4557 \\
$\epsilon_P^{(u)}$ & -2.1320 & -2.1254 & -2.1322 & -2.1320 & -2.1413 \\  \hline
\end{tabular}
\lae{tab1}
\end{table}

\begin{table}
\centering
\caption{Temperature and pressure jump coefficients for complete ($\sigma=1$)and non-complete ($\sigma \ne 1$) condensation and evaporation at the interface.}
\begin{tabular}{cccccc}\\ \hline
$\sigma$ & $\epsilon_T^{(T)}$ & $\epsilon_P^{(T)}$ & $\epsilon_T^{(u)}$ & $\epsilon_P^{(u)}$ &
$\epsilon_P^{(u)}$ \cite{Cip01}\\
\hline
0.1 & 1.9540 & 0.8404 &-0.4467 & -33.9467 & -34.0296\\
0.2 & 1.9540 & 0.8389 &-0.4468 & -16.3102 & -16.3050\\
0.4 & 1.9540 & 0.8381 &-0.4468 & -7.4489  & -7.4428\\
0.6 & 1.9540 & 0.8379 &-0.4468 & -4.4951 & -4.4887 \\
0.8 & 1.9540 & 0.8377 &-0.4468 & -3.0182 & -3.0116\\
1.0 & 1.9540 & 0.8376 &-0.4467 & -2.1320 & -2.1254\\ \hline
\end{tabular}
\lae{tab2}
\end{table}

\clearpage

\clearpage

\end{document}